%% file: main.tex
\documentclass[sigconf]{acmart}
\renewcommand\footnotetextcopyrightpermission[1]{}
\setcopyright{none} % No copyright notice required for submissions

\include{packages}

\graphicspath{{figures/}}

\begin{document}
%%%%%%%%%%%%%%%%%%%%%%%%%%%%%%%%%%%%%%%%%%%%%%%%%%%%%%%%%%%%%%%%%%%%%%%%%%%%%%%%%%%%%%

\fancyhf{} % Remove fancy page headers
\fancyhead{}
\fancyfoot[C]{\thepage}

\title{Lessons from HotStuff}

\author{Dahlia Malkhi}
\affiliation{%
  \institution{Chainlink Labs}
  \country{}
}
\author{Maofan Yin}
\affiliation{%
  \institution{Chainlink Labs\\Ava Labs, Inc.}
  \country{}
}

\date{\today}
\maketitle

\input{intro-material}
\input{nutshell}

\input{contribution}
\input{twophase}
\input{scaling}

\input{conclusion}

\section*{Acknowledgments}
We are grateful to multiple projects that adopted HotStuff and shared their insights and improvements with us, 
and 
for excellent input that helped improve this manuscript by Kartik Nayak, Mike Reiter, and Alberto Sonnino. 

%%%%%%%%%%%%%%%%%%%%%%%%%%%%%%%%%%%%%%%%%%%%%%%%%%%%%%%%%%%%%%%%%%%%%
%%%%%%%%%%%%%%%%%%%%%%%%%%%%%%%%%%%%%%%%%%%%%%%%%%%%%%%%%%%%%%%%%%%%%

%%%%%%%%%%%%%%%%%%%%%%%%%%%%%%%%%%%%%%%%%%%%%%%%%%%%%%%%%%%%%%%%%%%%%%%%%%%%%%%%%%%%%%
\bibliographystyle{plain}
\bibliography{bibliography}
%%%%%%%%%%%%%%%%%%%%%%%%%%%%%%%%%%%%%%%%%%%%%%%%%%%%%%%%%%%%%%%%%%%%%%%%%%%%%%%%%%%%%%

%%%%%%%%%%%%%%%%%%%%%%%%%%%%%%%%%%%%%%%%%%%%%%%%%%%%%%%%%%%%%%%%%%%%%%%%%%%%%%%%%%%%%%
\end{document}

%% file: packages.tex
\usepackage{xspace}
\usepackage{balance}

\usepackage[symbol]{footmisc}
\usepackage{filecontents}
\usepackage{graphicx}
\usepackage{hhline}
\usepackage{xspace}
\usepackage{balance}
\usepackage{ulem}
\usepackage{url}

\usepackage{breakurl}

\usepackage{titlesec}
\usepackage{colortbl}
\usepackage{multirow}
\usepackage{multicol}

 \usepackage{tikz}
\usepackage{amsmath}
\usepackage{enumerate}
\usepackage{enumitem}

\usepackage{algorithmicx}
\usepackage{algorithm}
\usepackage{algpseudocode}
\usepackage{varwidth}

\usepackage{amsthm}
\usepackage{textcomp}
\usepackage{gensymb}
\usepackage{makecell}
\usepackage{tabularx, booktabs}
\usepackage{array}
\usepackage{times}
\usepackage{fullpage}
\usepackage{dsfont}
\usepackage{epsfig}
\usepackage{microtype}
\usepackage{subcaption}
\usepackage{mathtools}
\usepackage{cleveref}
\usepackage{todonotes}
\usepackage{bm}

\newcommand{\mycomment}[1]{}

\newcommand{\nodes}{{nodes}\xspace}
\newcommand{\Nodes}{{Nodes}\xspace}

\newcommand{\numfaults}{\ensuremath{f}\xspace}

\newcommand{\event}[3]{
    \ifthenelse
    {\equal{#3}{}}
    {\ap{#1.\textrm{#2}}}
    {\ap{#1.\textrm{#2} \mid #3}}
}

\algnewcommand\Instance[2]{\State #1, \textbf{instance} #2}
\algnewcommand\InstanceSystem[3]{\State #1, \textbf{instance} #2, \textbf{system} #3}
\algnewcommand\Trigger[3]{\State \textbf{trigger} $\event{#1}{#2}{#3}$}
\algnewcommand\Schedule[1]{\State \textbf{schedule} $#1$}
\algnewcommand\Cancel[1]{\State \textbf{cancel} $#1$}
\algnewcommand\Broadcast[1]{\State \textbf{broadcast} $#1$}
\algnewcommand\Import[1]{\State \textbf{import} $#1$}

\algnewcommand\Not{\textbf{ not }}
\algnewcommand\AndT{\textbf{ and }}
\algnewcommand\OrT{\textbf{ or }}
\algnewcommand\In{\textbf{ in }}

\algsetblockdefx[Implements]{Implements}{EndImplements}{}{}{\textbf{Implements:}}{}
\algsetblockdefx[Uses]{Uses}{EndUses}{}{}{\textbf{Uses:}}{}
\algsetblockdefx[Init]{Init}{EndInit}{}{}{\textbf{Init:}}{}
\algsetblockdefx[Parameters]{Parameters}{EndParameters}{}{}{\textbf{Parameters:}}{}
\algsetblockdefx[Struct]{Struct}{EndStruct}{}{}[1]{\textbf{Struct} $#1$ \textbf{contains}}{}

\algsetblockdefx[Upon]{Upon}{EndUpon}{}{}[1]{\textbf{upon} $#1$ \textbf{do}}{}
\algsetblockdefx[UponExists]{UponExists}{EndUponExists}{}{}[2]{\textbf{upon exists} $#1$ \textbf{such that} $#2$ \textbf{do}}{}
\algsetblockdefx[UponCondition]{UponCondition}{EndUponCondition}{}{}[1]{\textbf{upon} $#1$ \textbf{do}}{}
\algsetblockdefx[UponReceiving]{UponReceiving}{EndUponReceiving}{}{}[1]{\textbf{upon receiving} $#1$ \textbf{do}}{}
\algsetblockdefx[UponEvent]{UponEvent}{EndUponEvent}{}{}[1]{\textbf{upon event} $#1$ \textbf{do}}{}

\algsetblockdefx[Process]{Process}{EndProcess}{}{}[2]{\textbf{process} \emph{#1}($#2$) \textbf{:}}{}
\algsetblockdefx[Function]{Function}{EndFunction}{}{}[2]{\textbf{function} \emph{#1}($#2$) \textbf{:}}{}
\algsetblockdefx[Procedure]{Procedure}{EndProcedure}{}{}[2]{\textbf{procedure} \emph{#1}($#2$) \textbf{is}}{}
\algsetblockdefx[FunctionNoArgs]{FunctionNoArgs}{EndFunctionNoArgs}{}{}[1]{\textbf{function} \emph{#1}() \textbf{:}}{}

\algdef{SE}[DoUntil]{DoUntil}{EndDoUntil}{\textbf{do}}[1]{\textbf{until} $#1$}
\algsetblockdefx[Struct]{Struct}{EndStruct}{}{}[1]{\textbf{Struct} $#1$ \textbf{contains}}{}
\algsetblockdefx[IfExists]{IfExists}{EndIfExists}{}{}[2]{\textbf{if exists} $#1$ \textbf{such that} $#2$ \textbf{then}}{\textbf{end if}}
\algsetblockdefx[While]{While}{EndWhile}{}{}[1]{\textbf{while} $#1$ \textbf{do}}{\textbf{end while}}
\algsetblockdefx[NTimes]{NTimes}{EndNTimes}{}{}[1]{\textbf{for} $#1$ \textbf{times do}}{\textbf{end for}}
\algsetblockdefx[For]{For}{EndFor}{}{}[3]{\textbf{for} $#1 \in #2..#3$ \textbf{do}}{\textbf{end for}}
\algsetblockdefx[ForAll]{ForAll}{EndForAll}{}{}[2]{\textbf{for} $#1 \in #2$ \textbf{do}}{\textbf{end for}}

%% file: intro-material.tex
\section{Introduction}

Every time you use a cloud service, there are servers behind the scenes keeping redundant copies of your data, solving the distributed \emph{consensus} problem to keep information available and consistent against the breakdown of some servers;
Every time you fly a modern airplane, there are extra sensors and avionics to keep it airborne, reaching a consensus on automated inputs to flight controls against malfunctioning components;
At the core of blockchains are systems that also solve consensus, to collectively maintain an immutable history of transactions against the worst type of failures, ``Byzantine'', orchestrated by rogue participants.

For four decades, experts in the field of distributed computing searched for optimal solutions to the classical Byzantine Fault Tolerant (BFT) consensus problem~\cite{LPS82}. Recently, a master thesis titled ``Consensus in the Age of Blockchains''~\cite{buchman2016tendermint}, which
was looking for a blockchain solution that developers can understand, changed the way we think about the problem. It led to the introduction of HotStuff~\cite{HotStuff-19}, the first practical solution (the meaning of being considered practical is defined below) with optimal communication complexity, that emerged 
as a new algorithmic foundation for the classical BFT consensus problem and a golden standard for 
blockchains.

This article will take you on a journey from the emergence of HotStuff to
lessons from it along two dimensions, foundational and applied.
The first part, Sections~\ref{sec:nutshell}-Section~\ref{sec:two-phase-hs},
underscores the theoretical advances HotStuff enabled, 
including several models in which HotStuff-based solutions closed problems which were open for decades.
This part finishes off with a surprising recent observation, HotStuff-2~\cite{cryptoeprint:2023/397},
demonstrating that it is possible to improve the original HotStuff latency by as much as 33\% without sacrificing any of its desirable properties (Section~\ref{sec:two-phase-hs}).
The second part, Section~\ref{sec:scaling},
 focuses on HotStuff performance in real life settings, where its simplicity drove adoption of HotStuff as the golden standard for blockchain design, and many variants and improvements built on top of it.
%
%Both parts of this document are meant to describe lessons drawn from HotStuff as well as disspell certain myths. 

\section{Preliminaries}
\label{sec:prelim}
%%%%%%%%%%%%%%%%%%%%%%%%%%%%%%%%%%%%%%%%%

\paragraph{The Problem.}
Briefly, in log replication, a group of hosts referred to as \nodes reach agreement on a growing sequence of bundled values called ``blocks''.
%There are various knobs and switches which may be tuned in a replication protocol.
For our purposes,
a solution is viewed as ``practical'' if it maintains consistency against any unforeseen network delays and advances at network speed, namely, as soon as a certain threshold of messages are received from participants.
This settings is known as \emph{partially-synchronous}.

More specifically, partially-synchronous BFT consensus replicates a log among $n=3\numfaults+1$ \nodes, $\numfaults$ of which are Byzantine. Byzantine nodes may collude and deviate from the specified protocol arbitrarily, though still with some common constraints (e.g., cannot have infinite computational power).
There is a known bound $\Delta$ on message transmission delays (neglecting message processing as marginal), such that after an unknown Global Stabilization Time (GST), all transmissions arrive within $\Delta$ bound to their destinations.

\Nodes output increasing log prefixes with the following guarantees:

\begin{description}
\item[Safety] At all times, for every pair of correct \nodes, the output log of one is a prefix of the other.

\item[Liveness] After GST, all non-faulty \nodes repeatedly output (growing) logs.
%decisions to extend the log by one position are repeatedly made by at least one correct validator.
\end{description}
 
We additionally desire to simultaneously achieve $O(n^2)$ worst-case communication, optimistically linear communication, an optimistically fast latency, and optimistic responsiveness. We define these properties more formally below.

\paragraph{Performance measures.}
Theoretical complexity measures are evaluated after GST, since no progress is guaranteed until then. There are two principal complexity measures: communication, measured in the number of bits sent over communication channels (by one node or in total); and latency, measured in units of network delays, maximal ($\Delta$) or actual ($\delta$).
We are interested in several aspects of these measures (communication and/or latency): expectation, optimistic, and worst-case.

Measures expressed as expectations are taken over protocol coin tosses, 
notably for electing ``leaders'' internally (see Section~\ref{sec:nutshell}). 
Optimistic performance measures are taken in faultless, synchronous executions. These measures also reflect
the protocol performance after a certain stabilization time following GST, but this analysis is left out of this short paper. 
Worst-case performance measures are taken against an unlucky cascade of $O(n)$ (leader) failures.

The desirable performance goals, which are derived from several known lower bounds, are as follows:

\begin{description}

\item[Latency.]
A solution has \emph{optimistic responsiveness} if optimistic latency is $O(\delta)$ per decision.
An $\Omega(n\Delta)$ worst-case latency is mandated by the Aguilera-Toueg bound~\cite{aguilera1999simple}.

\item[Communication.]
A solution is worst-case communication optimal if it incurs $O(n^2)$ communication cost~\cite{dolev1985bounds}.
The best communication cost to optimistically reach is $O(n)$ (the lower bound is trivial). 

\item[Load-Balance.]
A solution has \emph{load balance} if the same communication cost is incurred per party over a sequence of consensus decisions. Notably, this implies rotating leaders regularly.

\end{description}

It's worthy of noting that throughput is not a theoretical complexity measure.
We discuss the throughput of various systems in Section~\ref{sec:scaling}.

%% file: nutshell.tex
\section{Why HotStuff?}
\label{sec:nutshell}
%%%%%%%%%%%%%%%%%%%%%%%%%%%%%%%%%%%%%%%%%%%%%

\begin{figure*}[ht]
  \centering
  \includegraphics[width=\textwidth]{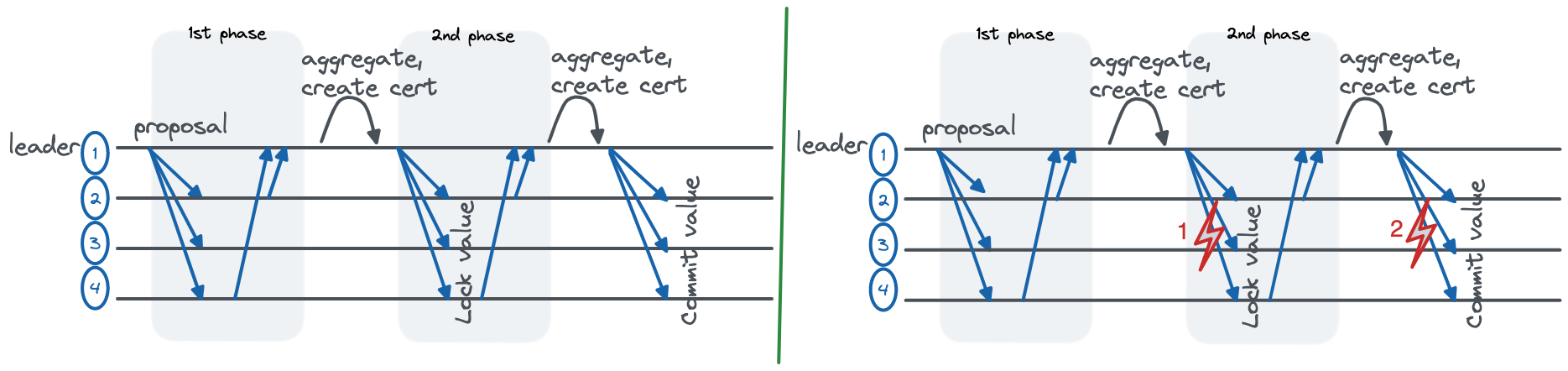}
  \caption{Two-step protocol, each step a linear secure broadcast (left). Possible failures during ratify step (right).}
  \label{fig:twostep}
\end{figure*}

In order to understand the improvement HotStuff introduces, let us consider a brief evolution of practical BFT solutions that led to it and the scaling properties they targeted.

\paragraph{View-by-View Recipe.}
PBFT~\cite{PBFT}, a landmark in BFT solutions introduced two decades ago, emphasizes optimistically low latency.
It established a view-by-view ``recipe'' that works as follows.
A view consists of two abstract steps. In the first step, a designated \emph{leader} attempts to \emph{reconcile} an output value, and in the second step, nodes \emph{ratify} if there is agreement and commit it. 
An advantage of this leader-based regime is that it is optimistically responsive (defined in Section~\ref{sec:prelim}),
that is, under synchronous faultless settings,
it does not need to wait for the maximal $\Delta$ delay, it instead incurs the actual network delay $\delta$.
Therefore, PBFT exhibits a desirable feature, responsiveness, during optimistic settings:

%Furthermore, a non-faulty leader will succeed in driving a commit decision by involving nodes in two voting steps.
%A voting step may take one network latency ($\delta$) in an all-to-all voting regime, or These two logical steps do not map one-to-one to latency in terms of $\delta$, because there are different voting regimes for PBFT variants, e.g., all-to-all or leader-driven. referred to as a \emph{two-phase view sub-protocol}.
%These two desirable features are captured below:

\begin{description}
%    \item[\textbf{F-0}] Two-phase view sub-protocol 
    \item[\textbf{F-1}] Optimistic responsiveness
\end{description}

\paragraph{Linear Secure Broadcast.}
PBFT employs a \emph{secure broadcast} building-block to disseminate a leader proposal. 
A secure broadcast provides a guarantee that non-faulty nodes deliver the same message from a sender, if any, and that messages from non-faulty leaders are reliably delivered. 
A second secure broadcast is used for assembling $2\numfaults+1$ votes to commit the proposal. 

PBFT’s original secure broadcast protocol is based on a protocol by Bracha~\cite{Bracha1987AsynchronousBA} and has quadratic communication complexity. 
Two pioneering works in the field, VBA~\cite{CachinKPS01} and Rampart~\cite{rampart-94}, which were later adopted in SBFT~\cite{SBFT-19}, 
employ signature aggregation for secure broadcast whose communication complexity is linear: A sender collects signatures on its message by a quorum of $2\numfaults+1$ out of $n=3\numfaults+1$, aggregates the signatures into a Quorum Certificate (QC) and disseminates the QC. 
%Each ``block'' in Figure ~\ref{fig:threestep} depicts a single linear secure broadcast.

Replacing PBFT's secure broadcast with a linear variant yields a two-step protocol depicted in
Figure~\ref{fig:twostep}(left), each step a linear secure broadcast, and achieves the following feature:

\begin{description}
    \item[\textbf{F-2}] Optimistic communication linearity
\end{description}

%We will add validity to step 1 after we discuss the implementation of the commit-adopt step.

\paragraph{View-Change with Quadratic Complexity.}

If a leader fails or the network stalls (before GST) during the ratify step, as depicted in Figure~\ref{fig:twostep}(right),
a new leader needs to check if any value is locked by a node from a previous view, and ratify it.

The ratify step in all the above protocols uses a lock-commit paradigm (aka commit-adopt~\cite{gafni98}),
where sufficiently many nodes are locked before any node can commit. 
If a new leader does not learn of any locked value, it can make a different proposal.
However, if it turns out that some nodes are locked on another value, they nevertheless need to vote for the (safe) new proposal to allow progress.
Consequently, in PBFT, 
new leader must \emph{prove} that $2\numfaults+1$ nodes did not vote to commit a different proposal. 
This approach for justifying a new leader proposal after a view-change is the foundation of all protocols in the PBFT family, 
including FaB~\cite{FaB-06}, Zyzzyva~\cite{zyzzyva-07}, Aardvark~\cite{Aardvark}, SBFT~\cite{SBFT-19}, 
and most former protocols in the two-phase HotStuff family~\cite{jalalzai2020fast,diembft-v4,gelashvili2022jolteon,mscfcl,wendy} except HotStuff-2, which we will get to later. 
Unfortunately, this justification proof is complex to code and incurs quadratic communication complexity. 

\paragraph{Simplified View-Change without Responsiveness.}
Tendermint~\cite{buchman2016tendermint} introduced a simpler view-change sub-protocol than PBFT, later adopted in Casper~\cite{casper}.
%that has linear complexity.
A new leader proposal simply hinges on the latest locked value (the highest block receiving a QC) the leader knows.
In fact, this simplification turns a new leader sub-protocol identical to a steady leader sub-protocol.
That is, in Tendermint there is no explicit view-change sub-protocol.
This provides another crucial tenet for blockchains: rotating leaders routinely, 
balancing participation and control among all nodes, as captured by the following feature:

\begin{description}
    \item[\textbf{F-3}] Balanced communication load over sequences of decisions 
\end{description}

However, to guarantee that a leader obtains the latest locked value in the system,
a leader in Tendermint has to wait for the maximal network delay $\Delta$. 
Hence, it does not satisfy optimistic responsiveness (F-1), 
namely, each view sub-protocol incurs an explicit delay for the maximal network latency. 
Moreover, the view sub-protocol is simpler but has the same complexity as PBFT, $O(n^2)$.
Nevertheless, Tendermint provided a crucial step in simplifying the view-change that is harnessed in HotStuff. 

\begin{figure}[ht]
  \centering
  \includegraphics[width=\columnwidth]{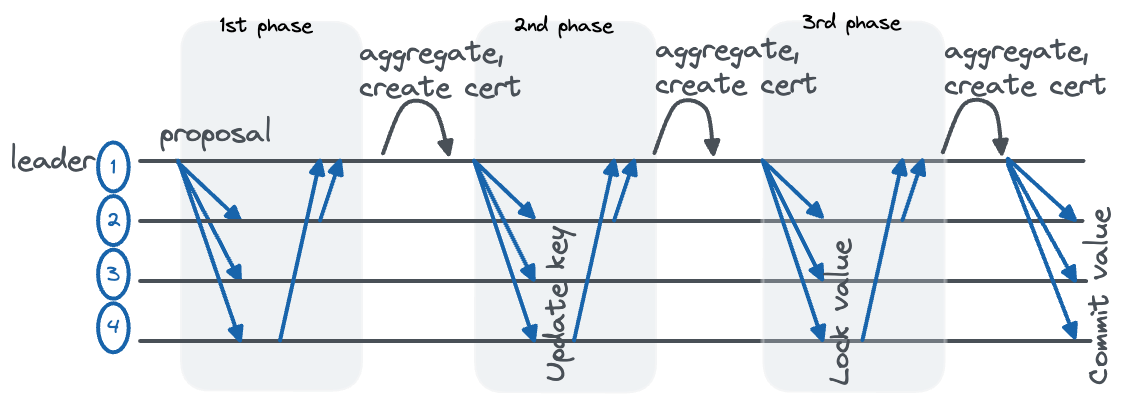}
  \caption{HotStuff three-step protocol.}
  \label{fig:threestep}
\end{figure}

\paragraph{Linear, Simple View-Change with Responsiveness.}

HotStuff~\cite{HotStuff-19} harnesses and enhances the simple Tendermint view-change in the following manner:

First, it removes the need for each view-change to delay, thereby satisfying F-1 in addition to F-2 and F-3.
This is achieved by employing three consecutive secure broadcasts, instead of two, to form a decision, as depicted in Figure~\ref{fig:threestep}. 
The first broadcast forms a QC guaranteeing the uniqueness of a leader proposal; the second provides $2\numfaults+1$ nodes with a copy of the QC (referred to as ``key'') to pass to the next leader, before any node can become locked or commit a value; the third confirms that $2\numfaults+1$ have a key and commits the value. 

\begin{figure}[ht]
  \centering
  \includegraphics[width=\columnwidth]{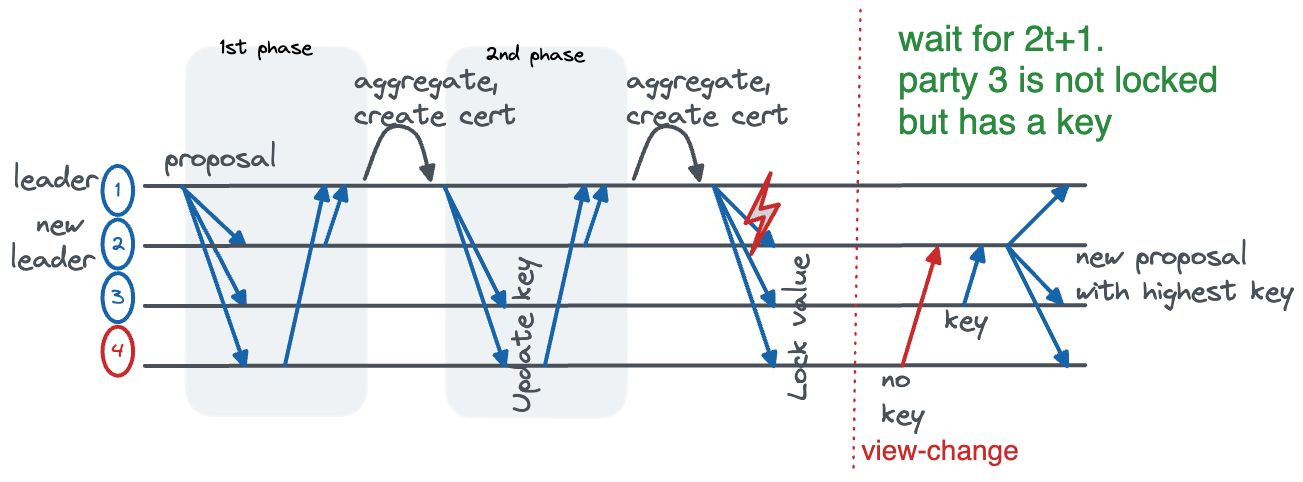}
  \caption{HotStuff view-change scenario.}
  \label{fig:threestep2}
\end{figure}

In a way, HotStuff spreads the lock-commit ratification step over two linear secure broadcasts.
The extra phase guarantees that if any party is locked on a leader proposal, then $2\numfaults+1$
already obtained a key corresponding to this lock. 
Correspondingly, the next leader would learn about the latest lock even if $\numfaults$ are Byzantine. 
In Figure~\ref{fig:threestep2}, the new leader (party 2) obtains the key from party 3 (Byzantine party 4 may not send its key), despite party 3 itself not reaching the lock stage.

Second, HotStuff employs linear secure broadcast to spread a leader proposal, making the view-change linear.

Third, a view-change sub-protocol must additionally address
view-synchronization, referred to as a \emph{Pacemaker} in~\cite{HotStuff-19}. 
A Pacemaker coordinates for nodes to enter
the next view roughly at the same time as the leader in order to guarantee progress.
%Former Pacemaker solutions used to incurred quadratic complexity, since it was anyway subsumed by the view change complexity.
RareSync~\cite{Civit2022ByzantineCI} and Lewis-Pye~\cite{LewisPye2022QuadraticWM} demonstrate a Pacemaker, which was mentioned only at a high-level in HotStuff~\cite{HotStuff-19}, that has worst-case $O(n^2)$ communication complexity.

Jointly, these enhancement achieve the following feature:

\begin{itemize}
    \item[\textbf{F-4}] Worst-case communication optimality
\end{itemize}

In summary, all the mentioned desirable performance properties (F-1,2,3,4) are simultaneously achieved by HotStuff with an optimal Pacemaker.
It is worth noting that the HotStuff family of protocols suffers an extra phase within the view sub-protocol compared with PBFT and Tendermint. We will come back to this in Section~\ref{sec:two-phase-hs}.

%% file: contribution.tex
\section{HotStuff Key Contributions}
%We are now ready to explain key contributions in HotStuff: chaining and linearity.

%%%%%%%%%%%%%%%%%%%%%%%%%%%%%%%%%%%%%%%%%%%%%%%%%%%%%%%%%%%%%%%%%%%%%
\subsection{Pipelining}

An important property stemming from the simplified leader replacement protocol is that all three secure broadcast steps of HotStuff are essentially identical.
This led to a key contribution introduced in HotStuff, namely, pipelining the protocol over a chain of blocks, each block embodying one step of the protocol. Furthermore, each block can be proposed by a different leader. 

\begin{figure}[h]
  \centering
  \includegraphics[width=\columnwidth-1in]{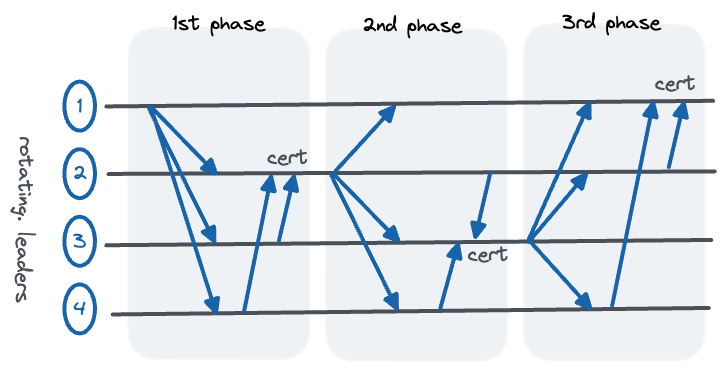}
  \caption{Pipelining.}
  \label{fig:pipelined}
\end{figure}

Each block in a pipeline is constructed by a leader proposal in one view and becomes certified via secure broadcast.
The next view proposes a block which is chained to the previous one, constituting the second step of the first proposal, and simultaneously, the first step of a new proposal. And so on.
Figure ~\ref{fig:pipelined} depicts a pipeline of three blocks, the first of which becomes committed.

\begin{figure}[h]
  \centering
  \includegraphics[width=\columnwidth]{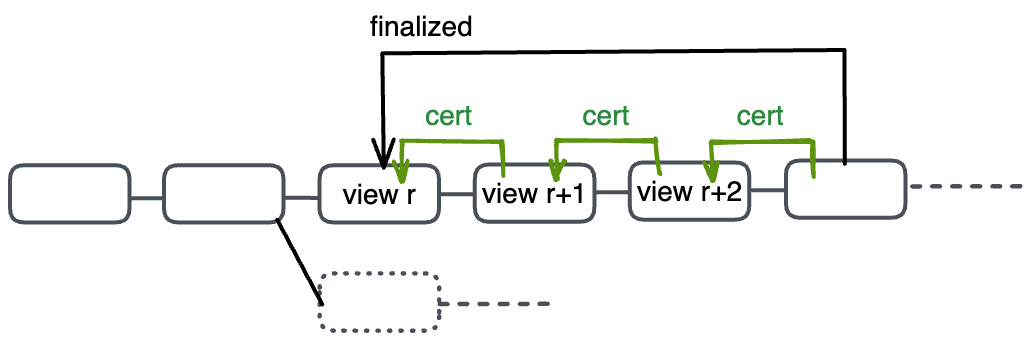}
  \caption{Three-chain commit rule.}
  \label{fig:pipelined2}
\end{figure}

The most important outcome of HotStuff pipelining is that it is easy to understand how the protocol constructs a replicated chain of blocks. 
Figure ~\ref{fig:pipelined2} below provides an easy visual explanation of the HotStuff three-chain rule: whenever the depicted three-block pattern occurs, the head of the three-chain becomes committed.

Prior to HotStuff, log replication solutions reached consensus one position at a time via a multi-round protocol, in a notoriously sophisticated fashion~\cite{DBLP:journals/usenix-login/Mickens14,abraham2017revisiting}.
Contrarily, HotStuff is apparent and intuitive, one simply looks for an uninterrupted chain of blocks to identify a consensus decision. View changes that are necessary
to resume the replication are depicted by forks from the main branch. 
% DM: Omit this sentence. It seems a bit overstated. HotStuff still has to handle message events, handle votes differntly when you are leader, etc. Also, it is subsumed by the following sentence.
%There is no need to reason about asynchronously delivered events and juggle around quite a few protocol state variables.

HotStuff also manages to encode almost all of its protocol state
into the data (i.e., blocks) it replicates, reducing to just two types of
messages: block proposals and votes. The execution of the protocol and final commitment
are made solely by checking the immutable chain fabrication that implicitly represents
a consistent causal ordering among all messages. This inspired
``zero-cost'' consensus protocols~\cite{danezis2022narwhal,spiegelman2022bullshark} that also use blocks to vote and thus entirely operate
upon the data it replicates without extra message exchange.

%the longest-fork-wins (LFW) protocol at the core of Bitcoin. 
%In LFW, each block is itself a full consensus decision embodying a proposal and votes. There are no additional protocol messages, hence LFW is extremely intuitive. However, LFW has drawbacks, including energy consumption and lack of immediate finality. 

%HotStuff combines the benefits of both classical BFT and LFW:
%
%\begin{itemize}
%\item Once reached, a consensus decision commits a block to the global sequence with immediate finality. 
%\item The procedure for finalizing a block involves additional blocks which help drive the commit decision. These blocks, themselves, become finalized in a pipeline, hence they increase throughput. 
% \end{itemize}

%HotStuff was the first BFT protocol to bridge between the classical BFT world and blockchains.
%Chaining is the most important reason HotStuff was understood by developers and became popular in the blockchain arena. 
%Today, practically every blockchain consensus core starts with a chained structure. 
%A pedagogical framework concentrating on chaining was given in Streamlet~\cite{cryptoeprint:2020/088}.

%%%%%%%%%%%%%%%%%%%%%%%%%%%%%%%%%%%%%%%%%%%%%%%%%%%%%%%%%%%%%%%%%%%%%
\subsection{Linearity}

The linear fastpath and view-change subprotocols of HotStuff empowered several tight solutions to open challenges in the consensus arena.

\paragraph{Partially synchronous BA.}
The most direct tight results enabled by HotStuff are
RareSync~\cite{Civit2022ByzantineCI} and Lewis-Pye~\cite{LewisPye2022QuadraticWM}, the first optimal partially-synchronous Byzantine agreement solutions, 
whose worst-case communication complexity is $O(n^2)$ with $O(n\Delta)$ latency (recall, worst-case complexities are taken after GST, against a cascade of $\numfaults$ actual leader failures).
Both solutions address HotStuff's view-synchronization black-box component, solving it with both expected and worst case $O(n^2)$ communication complexity.

\paragraph{Asynchronous BA.} VABA~\cite{VABA-19} is the first optimal solution to the long standing validated asynchronous Byzantine agreement problem\footnote{In a nutshell, validated agreement enforces an external validity predicate on decisions, rather than the theoretical Byzantine agreement problem formulation requiring all nodes to start with the same input.}
 whose communication complexity is $O(n^2)$. 

VABA invokes $n$ simultaneous HotStuff consensus instances, one per node acting as leader. 
After $n-\numfaults$ instances complete, VABA elects in retrospect one node as leader unpredictably and uniformly at random. 
It will either have reached a decision by this leader, or it orchestrates $n$ view-changes from it to the next wave of $n$ instances. 
Running $n$ simultaneous views and electing a random leader in restrospect has been suggested before in~\cite{KATZ200991}, but the HotStuff linear view-change enabled managing $n$ view-changes with overall complexity $O(n) \cdot n = O(n^2)$. 
 
\paragraph{Optimistically Asynchronous BA.}

Bolt-Dumbo~\cite{Lu2021BoltDumboTA} and Jolteon and Ditto\cite{gelashvili2022jolteon} demonstrate an optimistically asynchronous Byzantine agreement,
% in~\cite{..}, because it is more than a single citation
a problem pioneered in~\cite{10.1007/11523468_17,Ramasamy-parsimonious-05}.
They use a two-phase variant of HotStuff as an optimistically linear path, for the case of a non-faulty leader and partial synchrony settings.
They employ a quadratic asynchronous protocol as fallback upon a leader failure, thereby providing resilience against asynchrony.

% I'm removing these.
% I don't think we'll have room for them,
% and hopefully the performance section contains enough of the context about the DAG and so forth.
% \subsection{Near ``Zero-Cost'' Consensus}
% \TY{the title was originally ``Protocol on a Graph'', but maybe a more attractive title is better\ldots}
% \TY{To be filled with discussion on how most of the protocol state is embedded in the ``chain'' (which is in fact, a tree-shaped graph) as the data structure to be disseminated and why that helps simplify protocol designs and may have inspired other chain-based/DAG-based protocols such as the ``zero-cost'' consensus in Narwhal-Tusk.}
% \paragraph{Implicit view change.}
% \paragraph{Consensus by checking condition.}

% ..........................
\subsection{The Pacemaker Module}
%\TY{We probably also need a ``Modularity'' for the Pacemaker. At least according to Zekun's words, he found the separation of liveness/performance from the safety very helpful in implementing Libra/DiemBFT (where our protocol shines engineering-wise, compared to other candidates).}

The \emph{Pacemaker} abstraction introduced by HotStuff captures the view synchronization challenge as a separate module in BFT consensus.
This modularity contributed further to HotStuff developer friendliness. 
Additionally, the formulation of the Pacemaker as a problem in itself
has sparked interest, leading to several advances.

Briefly, a Pacemaker solves the Byzantine view synchronization problem, where a group of processes enter/leave views until they reach a view with a non-faulty leader and spend sufficient overlapping time in the view for the leader to drive a consensus decision. 
Before HotStuff, BFT solutions for the partial synchrony settings required quadratic communication complexity per view-change, hence no one cared if coordinating entering/leaving a view also incurs quadratic communication. Linearity has shifted the challenge to developing a Pacemaker with low communication.

Cogsworth~\cite{Naor2019CogsworthBV} and a protocol by Naor and Keidar (NK)~\cite{Naor2020ExpectedLR} demonstrate Pacemakers with expected linear communication complexity and worst case $O(n^3)$. 
Expected linearity is achieved via the following strategy.
When nodes want to move to the next view, they send a message only to the next view's leader. The leader collects the messages from the nodes, and once it receives enough messages, it combines them into a threshold signature and sends it to the nodes. This all-to-leader, leader-to-all communication pattern is similar to the one used in HotStuff;
the trick in Cogsworth/NK is utilizing $\numfaults+1$ consecutive leaders as fallback \emph{relayers}, staggering leaders one at a time—each after a (tunable) Pacemaker timeout, until there is progress. One of the relayers is non-faulty and will facilitate entering the next view.

Two recent works, 
RareSync\cite{Civit2022ByzantineCI} and Lewis-Pye (LP)~\cite{LewisPye2022QuadraticWM}, solve the view synchronization problem with both expected and worst case $O(n^2)$ communication complexity.
Both use a similar approach, which is remarkably simple and elegant. It bundles consecutive views into epochs, where each epoch consists of $\numfaults+1$ consecutive views. Nodes employ a Bracha-like all-to-all coordination protocol in the first view of each epoch, and then they advance through the rest of the views in the same epoch using timeouts if there is no progress in the underlying consensus protocol.
The downside of RareSync/LP is that the expected message complexity and latency are as bad as the worst case, hence the expected case performance is worse than previous solutions. 

It remains open and an active area of research to find view-synchronization solutions with both optimal worst case and expected/optimistic performance. 
Further discussion of view synchronization appears in~\cite{viewsync-post-22}.

%% file: twophase.tex
\section{Two-Phase HotStuff}
\label{sec:two-phase-hs}
%%%%%%%%%%%%%%%%%%%%%%%%%%%%%%%%%%%%%%%%%%%%%%

Since the introduction of HotStuff it remained an open challenge to achieve the desirable properties F-1,2,3,4 it encompasses with a two-phase view rather than a three-phase sub-protocol.
Recently, 
a two-phase HotStuff variant named HotStuff-2 was introduced in~\cite{cryptoeprint:2023/397} 
showing it is possible to simultaneously achieve all five desirable properties.
That is, it is possible to
solve partially-synchronous BFT and simultaneously achieve 
a two-phase commit sub-protocol within a view, 
optimistic responsiveness, 
optimistic communication linearity, 
balanced load across nodes, 
and
$O(n^2)$ worst-case communication. 
The main takeaway is that two phases are enough for BFT after all.

HotStuff-2 is remarkably simple, adding no substantial complexity to the original HotStuff protocol.
It builds on two secure broadcasts. The first step certifies with a QC uniqueness of a leader proposal. The second one is a lock-commit step for ratifying it.

\begin{figure*}[ht]
  \centering
  \includegraphics[width=\textwidth]{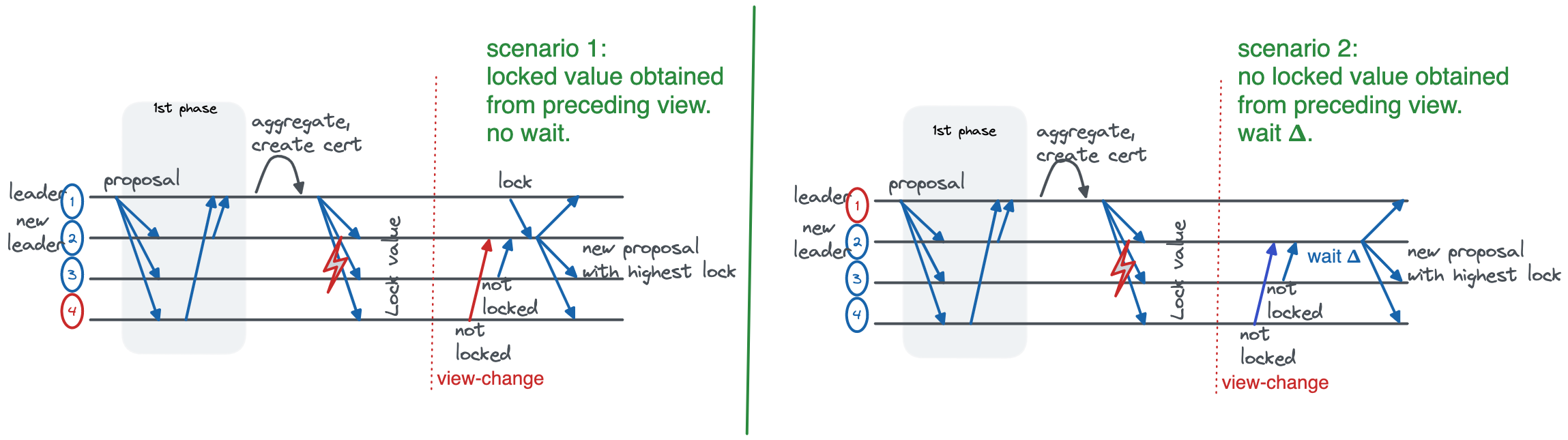}
  \caption{HotStuff-2: Some parties may not commit in a view but they become locked in it.\\\hspace{\textwidth} Case 1: the highest lock is obtained by the next view leader and it proceeds responsively (left).\\\hspace{\textwidth} Case 2: no honest party obtains a lock in a view, and the next view leader has to wait to propose in the next view (right).}
  \label{fig:hs2}
\end{figure*}

The key observation is that a new leader can choose between two options: 
If the leader obtains a QC from the preceding view, it \textbf{knows} that it has obtained the latest locked value that possibly exists in the system.
In this case, it proceeds with a proposal in a responsive manner.
Otherwise, the leader \textbf{knows} that a timer delay of $\Delta$ must have expired in the preceding view.
In that case, there is no responsiveness anyway, hence it waits an extra $\Delta$ to obtain the latest locked value in the system.
Figure~\ref{fig:hs2} depicts two possible HotStuff-2 view-change scenarios.

Prior to HotStuff-2, there has been a long line of HotStuff variants aiming to improve HotStuff's view regime to two phases. 
Fast HotStuff~\cite{jalalzai2020fast}, DiemBFT-v4~\cite{diembft-v4}, and Jolteon and Ditto~\cite{gelashvili2022jolteon}, provide two-phase view regimes but revert to a PBFT quadratic view-change (Ditto also adding resilience against asynchrony, as mentioned above).
Hence, they do not satisfy F-4, namely they incur $O(n^2)$ communication every time a leader is faulty. A fortiori, an unlucky cascade of faulty leaders incurs $O(n^3)$ communication.   
Wendy~\cite{wendy} and MSCFCL~\cite{mscfcl} also revert to a PBFT view-change with a leader proof to convince parties of a safe proposal, but focus on compressing the leader proof.
These schemes employ somewhat heavy hammers: Wendy introduces a novel signature scheme that works only when the gap between views that make progress is constant bounded and MSCFCL utilizes succinct arguments of knowledge whose complexity blows up quickly.
All of these advances are much more complex than HotStuff-2, whose suprising upshot is that none of them is necessary.

%% file: scaling.tex
\section{Scaling}
\label{sec:scaling}

Aside from theoretical considerations, practical consensus protocols also need
to be fast and scalable when it comes to actual implementation. Over the
past years, as we learned about HotStuff variants and studied subsequent
protocols, we extracted several insights about improving the performance of HotStuff and discovered some prevailing myths about scalability.

The main scalability challenge is the overhead of coordination among an increasing number of participating nodes 
and increasing network latencies among them.
The goal is to maintain high throughput and low latency.

\subsection{What is the ``Leader Bottleneck''}
% .............................................
The principal reason for using leader based consensus protocols in general, not just HotStuff,
which we've heard repeatedly from multiple blockchain projects,
is emphasis on low latency.
In particular, using the reconcile/ratify consensus recipe described in Section~\ref{sec:two-phase-hs}, 
a good leader can drive the reconciliation step in one network round-trip, 
and in just one more (logical) step,
agreement can be detected and committed.
However, one of the strongest weaknesses mentioned in the literature is the so-called
``leader bottleneck''. 

Specifically in HotStuff, the leader bottleneck is manifested in a pipeline of linear secure broadcasts.
In each instance in the pipeline,
first, a leader disseminates blocks and all other nodes are not communicating with one another, 
thereby the network bandwidth is underutilized;
second,
the leader collects signed messages from all nodes, validates (aggregates) the signatures, and updates its protocol state, while all other nodes are idle.
Linear secure broadcasts are invoked in a sequence, where each one has to wait for responses from
$2f+1$ nodes before it moves to the next step. This takes a full roundtrip
to and from the slowest node among the fastest $2/3$ of the network. In WAN settings 
with geo-distributed nodes, this almost always takes an order of
hundreds of milliseconds, including additional time spent verifying $2f+1$ votes. 

At first glance, it thus appears that low latency comes at the cost of bounded throughput.

We proceed to describe prevailing approaches for parallelizing work in order to saturate network and computational resources.
Some approaches are compatible with HotStuff and may be harnessed to increase its throughput; others hinge on new BFT consensus foundations.

% redundant , covered below. removing.
%Protocols can improve the performance by multiple leaders,
%but not logically for the same consensus instance. Having multiple ``leaders'' for the same instance cannot increase
%the performance because no matter how many leaders there are, they still collaboratively yield a single, consistent decision.
%On the contrary, more proposals will postpone the reconciliation process to reach a unanimous majority.
%To make more than a single decision in one instance (batching counts as one decision), we'll have to weaken the replication model so not all items need to be
%totally ordered (as in EPaxos), but that's another story.

\subsection{Saturating the Resources}
% .........................................

\paragraph{Parallel Computation.} 
A simple way to increase throughput is to offload networking and
computationally intensive tasks to \emph{workers}. Despite the sequential
skeleton of a consensus protocol, signature verification, ``mempool'' (a blockchain subsystem which buffers
transactions from clients and bundles them into blocks) synchronization,
and/or block dissemination, can be made parallel in between the key phases of the consensus.
For example,  we heard that from many real-life HotStuff systems that
the leader work is offloaded to a farm of CPUs or even to a local cluster of hosts,
each handling messages to/from other nodes and carrying verification in parallel. 

\paragraph{Large Blocks.} 
Another simple way to increase throughput and ameliorate the idle time caused by network
latency is to batch larger payloads per block. 
The key insight here is that the non-network time required to handle/process/execute a block grows linearly with block size, whereas network transmission time remains almost fixed, or grows very slowly.
This means that the utilization rate increases by larger blocks and throughput grows.
However, although this will increase throughput it will also increase latency.
Additionally, larger blocks do not scale throughput forever. 
In the limit, very large blocks increase latency to a point where further throughput may not be gained.
The long-version HotStuff paper~\cite{HotStuff-long} uses this technique, whose evaluation section shows the throughput saturates
at batching hundreds of transactions (``400 vs. 800'' curves). The sweet spot is adhoc to the specific
application and its transactions, varying across practical blockchain instantiations and their deployment.

\begin{figure}
  \centering
  \includegraphics[width=\columnwidth]{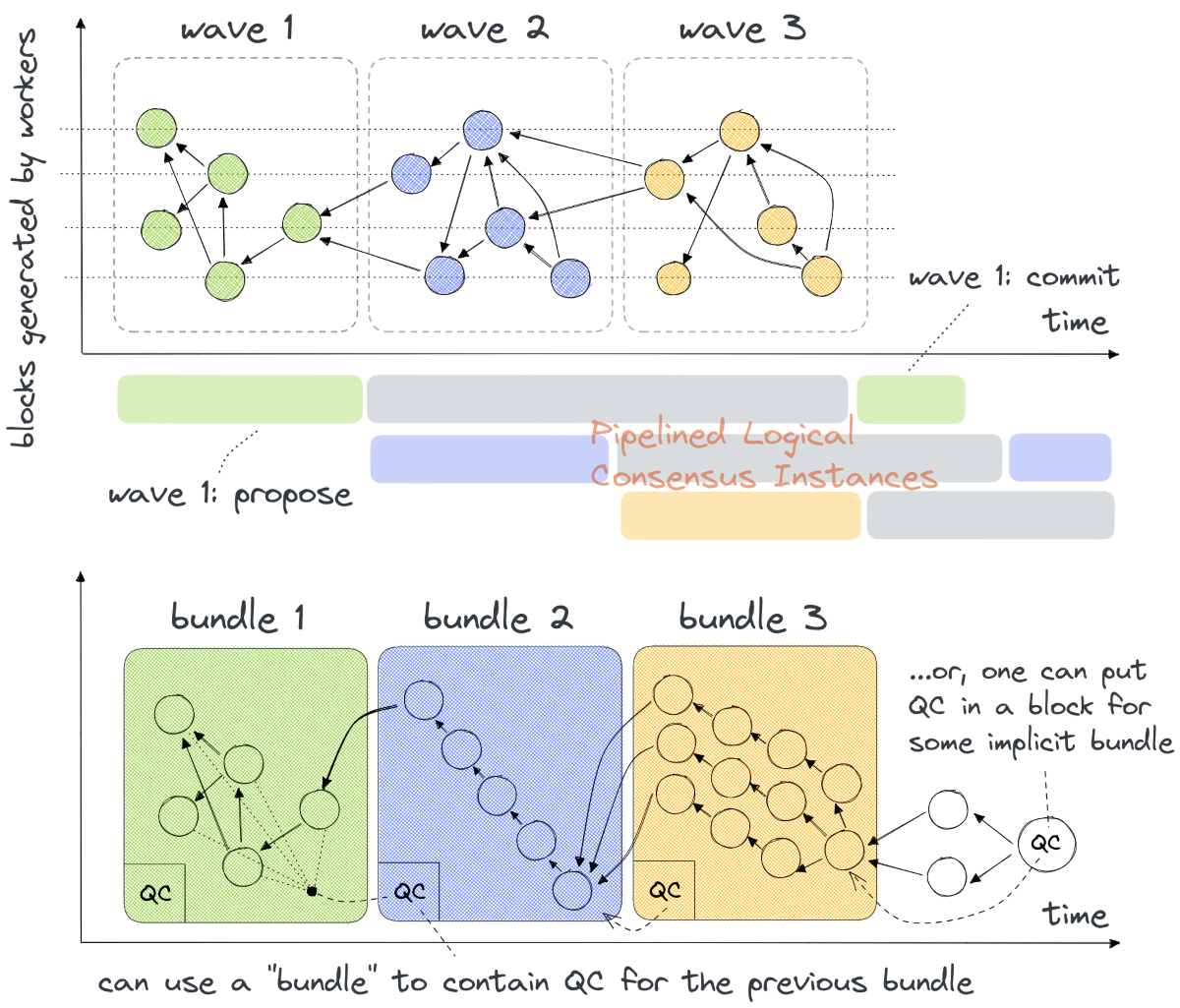}
  \caption{Driving waves/bundles of blocks.}
  \label{fig:block-wave}
\end{figure}

\paragraph{Block Waves.}
Recently, an approach built on a different consensus foundation has demonstrated excellent resource utilization by nodes working in parallel on proposing and parsing blocks and then driving a consensus decision on a \emph{wave} of blocks.
It is much more effective than batching because nodes can ``buffer'' blocks
collaboratively and then let a consensus decision commit the entire wave.
Moreover, information can continue spreading by nodes in the background while driving the next consensus decision,
so that even if consensus stalls, the network continues having utility.
More specifically, the idea is to let the entire network propose new blocks and organize
the blocks by a layered DAG where each layer corresponds to a logical phase of the consensus protocol~\cite{danezis2022narwhal,spiegelman2022bullshark}.
Then, by some deterministic graph traversal, the blocks of each wave could be pipelined to commit in a linear order, triggered by the key phases. The upper diagram of Figure~\ref{fig:block-wave} sketches this approach in terms of network scheduing. 

It is interesting to contrast the DAG approach with a ``smart mempool'' approach depicted in the bottom diagram.
The idea is that blocks can be proposed in parallel and disseminated to the mempool with causal relations.
Leaders can inject special blocks into the mempool,
forming ``bundles'' in their proposals and carrying QCs for previous bundles.
The main difference is that bundles can have free structures, as shown in the figure.
This is applicable to HotStuff and other chain-style protocols in general.

\paragraph{Concurrent Instances.}
Instead of carrying parallel work with the effort of a single leader,
one can run multiple consensus instances concurrently, aka a ``leaderless'' approach, as in~\cite{mirbft, smrmadesimple, LevAri2019FairLedgerAF, kang2023practical}. 
The core idea in these protocols is to partition the replicated chain (log) according to some rules (e.g., round-robin) into pre-designated slots.
All instances are performed in parallel by the nodes.
Of course a realistic scheme needs to be fault tolerant, hence it needs a mechanism to handle faulty instances.
This requires making a consensus decision, but the consensus method for this does not need to be high-throughput.
Like the wave approach, the main drawback of running concurrent instances is the increased latency to commit finality.

\paragraph{Sharding.}
Since consensus offers fault-tolerance by introducing redundancy,
scaling state-machine replication is fundamentally capped at the throughput of a single node. 
Therefore, the best scenario is that the replicas can perform as a cohort close to a single machine (the conceptual state machine being replicated) performance.
It is worth remarking that some blockchain projects scale-out via sharding~\cite{polkadot}, but this trades off fault tolerance, effectively reducing the global resilience down to the resilience of each single shard. Sharding is left out of scope from this short paper.

%% file: conclusion.tex
\subsection{Concluding Remarks}
% ...................................
%Further research is needed to quantify and systematically study the performance trade-offs of the above techniques.
We call on a systematic evaluation of the existing or emerging consensus systems, by clearly identifying the improvements
brought by any of the aforementioned techniques and their combinations.
Specifically, while the sequential logic
in a consensus instance is inevitable, one can offload as much as possible from the core logic so it is only left with lightweight
small state mutation that is just enough to bookkeep the protocol state, and then parallelize work on the rest.
Another important topic is separating data dissemination and availability from sequencing digests of the data. 
An additional issue is that end users usually do not directly participate in the consensus protocol, and thus
the mempool used for disseminating user requests could create fairness issues with respect to sequencing, known as Miner/Maximal Extractable Value (MEV).

However, common practice is to merely show full-system performance results and compare them against other full-systems, which are also complex. 
In our experience, various engineering optimizations and system considerations may have surprising performance gains that
have little to do with the fundamental consensus protocol. 
Moreover, common optimizations like batching and parallelizing message (signature) validation are applicable to many protocols. 
To avoid making apple-to-orange comparisons, the scientific community would benefit from a systematic, ingredient-by-ingredient study of performance.
Improving throughput, for example, affects the latency and it would be useful to know where it crosses a prohibitive point.
Careful engineering is another point which would be beneficial to isolate.

Ultimately, to arrive at a high-performance, carefully engineered system, requires employing multiple techniques to saturate both the network and
computational resources as much as possible. 

On the foundational side, additional effort in needed to improve Pacemakers: the holy grail is a Pacemaker with expected linear communication, 
worst-case quadratic communication, and only $O(\Delta)$ delay per leader failure. 
The introduction of HotStuff-2 opens a door for a new generation of protocols. 
For example, it would be interesting to explore merging methods that were previously introduced to improve latency in HotStuff (e.g., ~\cite{jalalzai2020fast,gelashvili2022jolteon,wendy}) into HotStuff-2. 
Another potential direction would be exploring if HotStuff-2 brings new insights or improvements in other fault models,
e.g., in Momose-Ren~\cite{momose-ren22} where the core structure of HotStuff is adapted to the Sleepy model of Pass and Shi~\cite{sleepy17}.